# Spin-selective reactions of radical pairs act as quantum measurements

J. A. Jones[1,2] and P. J. Hore[3,*]

[1]Centre for Quantum Computation, Clarendon Laboratory, University of Oxford, Parks Road, Oxford OX1 3PU, United Kingdom

[2]Centre for Advanced ESR, University of Oxford, South Parks Road, Oxford OX1 3QR, United Kingdom

[3]Department of Chemistry, University of Oxford, Physical and Theoretical Chemistry Laboratory, South Parks Road, Oxford OX1 3QZ, UK

*Corresponding author. Fax: +44 (0)1865 275410.
E-mail address: peter.hore@chem.ox.ac.uk (P.J. Hore).

## Abstract

Since the 1970s, spin-selective reactions of radical pairs have been modelled theoretically by adding phenomenological rate equations to the quantum mechanical equation of motion of the radical pair spin density matrix. Here, using a quantum measurement approach, we derive an alternative set of rate expressions which predict a faster decay of coherent superpositions of the singlet and triplet radical pair states. The difference between the two results, however, is not dramatic and would probably be difficult to distinguish experimentally from decoherence arising from other sources.

# 1. Introduction

Conservation of spin angular momentum is an important determinant of the reactivity of paramagnetic molecules [1,2]. For example, when two diffusing radicals each with electron spin $S$ = ½ meet at random in solution the encounter pair may have overall spin $S$ = 0 (a singlet state) or $S$ = 1 (a triplet state). Typically, the radicals react (recombine) to form diamagnetic products ($S$ = 0) but can only do so from the singlet state. Radicals that encounter in a triplet state are either unreactive towards one another (and so diffuse apart) or react to form a triplet product ($S$ = 1). The chemical reactions that create radical pairs are also spin-conserving. For example, a bimolecular electron transfer reaction between an excited triplet electron donor and a closed shell acceptor would produce a pair of ion radicals in a triplet state. However, singlet and triplet are rarely eigenstates of the spin Hamiltonian, with the consequence that the spin state of the radical pair evolves coherently at frequencies and with amplitudes determined by the internal and external magnetic interactions of the electron spins [3]. Simultaneously, the singlet and triplet radical pairs react at different rates to form distinct reaction products. These processes are summarized in Figure 1.

The interactions that determine the coherent singlet-triplet interconversion of a radical pair include Zeeman, hyperfine, exchange and dipolar interactions of the electron spins with applied magnetic fields, with nuclear spins and with each other. The resulting spin evolution typically falls in the $10^7$-$10^9$ Hz range and is often significantly faster than the electron spin relaxation (typically < $10^7$ $s^{-1}$) that results from time-dependent local magnetic fields. In many cases, there is therefore ample time for weak magnetic interactions to influence the spin dynamics and consequently the probabilities of reaction via the singlet or the triplet channel. Such effects form the basis of the sensitivity of radical pair reactions to applied magnetic fields [1,4-6], and are the origin of chemically induced electron and nuclear spin polarizations [7] and the magnetic isotope effect [8]. These phenomena have collectively come to be known by the name Spin Chemistry.

A proper account of spin chemical effects requires a quantum mechanical description of the coherent spin dynamics combined with a kinetic treatment of the spin-selective reactivity. Since the earliest days of Spin Chemistry, this has almost exclusively been done by augmenting the Liouville-von Neumann equation of motion for the spin density matrix $\hat{\rho}(t)$ with phenomenological rate equations for the disappearance of singlet and triplet radical pairs, with first order rate constants $k_{\mathrm{S}}$ and $k_{\mathrm{T}}$ [9-14]:

$$\begin{aligned} \frac{\mathrm{d}\rho_{\mathrm{SS}}}{\mathrm{d}t} &= -k_{\mathrm{S}}\rho_{\mathrm{SS}}; & \frac{\mathrm{d}\rho_{\mathrm{TT}}}{\mathrm{d}t} &= -k_{\mathrm{T}}\rho_{\mathrm{TT}} \\ \frac{\mathrm{d}\rho_{\mathrm{ST}}}{\mathrm{d}t} &= -\tfrac{1}{2}(k_{\mathrm{S}}+k_{\mathrm{T}})\rho_{\mathrm{ST}}; & \frac{\mathrm{d}\rho_{\mathrm{TS}}}{\mathrm{d}t} &= -\tfrac{1}{2}(k_{\mathrm{S}}+k_{\mathrm{T}})\rho_{\mathrm{TS}} \end{aligned} \tag{1}$$

where $\rho_{jk} = \langle j|\hat{\rho}(t)|k\rangle$, using the minimal basis: $|\mathrm{S}\rangle$ ($S$ = 0, $M_S$ = 0) and $|\mathrm{T}\rangle$ ($S$ = 1, $M_S$ = 0). According to Equation (1), the diagonal elements of $\hat{\rho}(t)$ in this basis are damped exponentially at rates $k_{\mathrm{S}}$ or $k_{\mathrm{T}}$ while the off-diagonal elements decay with the mean rate constant, $\tfrac{1}{2}(k_{\mathrm{S}}+k_{\mathrm{T}})$.

In this Letter, we re-examine the phenomenological approach from the standpoint of quantum measurement theory. Our attention was drawn to this matter by recent claims [15] that the spin dynamics of an experimental model system [16] mimicking the proposed avian radical pair compass [17] cannot be explained in terms of the traditional approach outlined above.

## 2. Phenomenological approach

The conventional approach for calculating the effects of spin-state selective chemical reactions on the spin dynamics of radical pairs is described by Haberkorn [12]; although he only considered reactions from the singlet state, his treatment can easily be extended to include parallel singlet and triplet selective reactions (see Figure 1). The equation of motion of the radical pair density operator $\hat{\rho}(t)$ in Hilbert space is

$$\frac{\mathrm{d}\hat{\rho}(t)}{\mathrm{d}t} = -\mathrm{i}\left[\hat{H}, \hat{\rho}(t)\right] - \tfrac{1}{2}k_\mathrm{S}\left(\hat{\rho}(t)\hat{Q}_\mathrm{S} + \hat{Q}_\mathrm{S}\hat{\rho}(t)\right) - \tfrac{1}{2}k_\mathrm{T}\left(\hat{\rho}(t)\hat{Q}_\mathrm{T} + \hat{Q}_\mathrm{T}\hat{\rho}(t)\right) \quad (2)$$

where $\hat{H}$ is the spin Hamiltonian, and $\hat{Q}_\mathrm{S}$ and $\hat{Q}_\mathrm{T}$ are the singlet and triplet projection operators. $\hat{Q}_\mathrm{S}$ and $\hat{Q}_\mathrm{T}$ have all the usual properties of projectors [18], in particular that $\hat{Q}_i^{\,2} = \hat{Q}_i$ and $\hat{Q}_\mathrm{S} + \hat{Q}_\mathrm{T} = \hat{E}$, where $\hat{E}$ is the identity operator. The first term on the right hand side of Equation (2) is the usual Liouville-von Neumann commutator for the coherent spin dynamics. In the $\{\mathrm{S}, \mathrm{T}\}$ basis, the second and third terms result in the kinetics of Equation (1).

It is more convenient to solve this equation in Liouville space where it becomes

$$\frac{\mathrm{d}}{\mathrm{d}t}\vec{\rho}(t) = -\hat{\hat{V}}\vec{\rho}(t); \qquad \hat{\hat{V}} = \mathrm{i}\hat{\hat{H}} + \tfrac{1}{2}k_\mathrm{S}\hat{\hat{Q}}_\mathrm{S}^{+} + \tfrac{1}{2}k_\mathrm{T}\hat{\hat{Q}}_\mathrm{T}^{+} \quad (3)$$

where the (anti)-commutator superoperators are defined by

$$\hat{\hat{A}}^{\pm} = \hat{A} \otimes \hat{E} \pm \hat{E} \otimes \hat{A}^\mathrm{T} \quad (4)$$

and $\hat{\hat{H}} \equiv \hat{\hat{H}}^{-}$. Integration of Equation (3) gives the formal solution

$$\vec{\rho}(t) = \exp\left(-\hat{\hat{V}}t\right)\vec{\rho}(0) \quad (5)$$

It is straightforward to evaluate this in the simple case $\hat{H} = 0$ as $\hat{\hat{V}}$ is then diagonal in the $\{|\mathrm{S}\rangle\langle\mathrm{S}|,\ |\mathrm{S}\rangle\langle\mathrm{T}|,\ |\mathrm{T}\rangle\langle\mathrm{S}|,\ |\mathrm{T}\rangle\langle\mathrm{T}|\}$ basis. Explicitly,

$$\exp\left(-\hat{\hat{V}}t\right) = \begin{pmatrix} \exp(-k_\mathrm{S}t) & 0 & 0 & 0 \\ 0 & \exp(-\tfrac{1}{2}[k_\mathrm{S}+k_\mathrm{T}]t) & 0 & 0 \\ 0 & 0 & \exp(-\tfrac{1}{2}[k_\mathrm{S}+k_\mathrm{T}]t) & 0 \\ 0 & 0 & 0 & \exp(-k_\mathrm{T}t) \end{pmatrix} \qquad (6)$$

indicating that the singlet and triplet populations will decay at rates $k_\mathrm{S}$ and $k_\mathrm{T}$ respectively, exactly as expected from a naïve treatment, while the off-diagonal superposition terms ($\rho_\mathrm{ST}$ and $\rho_\mathrm{TS}$) decay at the average of these two rates (as in Equation (1)).

The solution of this equation is not a proper density matrix (except at $t = 0$) as the evolution described by Equations (5) and (6) does not preserve the trace. This is unsurprising and simply corresponds to the disappearance of radical pairs to form reaction products. It can be circumvented by enlarging the Hilbert space to include the products. This can be done by including explicit chemical reaction terms [13,19,20], but it is simpler to note that the rate of appearance of singlet product is equal to the rate of decay of the singlet radical pair population, and equivalently for the triplet, so that since

$$\left\langle \hat{Q}_\mathrm{S} \right\rangle(t) = \mathrm{Tr}\left\{ \hat{Q}_\mathrm{S}\hat{\rho}(t) \right\} \quad \text{and} \quad \left\langle \hat{Q}_\mathrm{T} \right\rangle(t) = \mathrm{Tr}\left\{ \hat{Q}_\mathrm{T}\hat{\rho}(t) \right\} \qquad (7)$$

are the fractional singlet and triplet radical-pair populations then

$$k_\mathrm{S}\int_0^t \left\langle \hat{Q}_\mathrm{S} \right\rangle(t')\,\mathrm{d}t' \quad \text{and} \quad k_\mathrm{T}\int_0^t \left\langle \hat{Q}_\mathrm{T} \right\rangle(t')\,\mathrm{d}t' \qquad (8)$$

are the fractional product populations, and the sum of these four quantities is unity for all $t$ as desired. This approach implicitly assumes that coherent superpositions of radical-pair and product states cannot occur, but this is reasonable as the decoherence time of such superpositions will be very short.

## 3. A quantum measurement approach

Having described the conventional approach we now derive an equation of motion for the radical pair based on the theory of quantum measurements; the result differs significantly from Equations (3), (5) and (6). Other aspects of quantum control and entanglement in radical pairs have recently been explored [20,21].

Non-unitary evolutions, such as chemical reactions, are conveniently described using the operator-sum approach [18]

$$\hat{\rho}(t+\mathrm{d}t) = \sum_k p_k \hat{A}_k(\mathrm{d}t)\,\hat{\rho}(t)\,\hat{A}_k^\dagger(\mathrm{d}t) \tag{9}$$

where the operator $\hat{A}_k$ is applied with probability $p_k$, and the overall process is trace-preserving if

$$\sum_k p_k \hat{A}_k^\dagger(\mathrm{d}t)\,\hat{A}_k(\mathrm{d}t) = \hat{E}\,. \tag{10}$$

For our spin-sensitive reactions, during the interval ($t$, $t$ + d$t$) a fraction $k_\mathrm{S}\mathrm{d}t$ of radical pairs potentially undergoes the singlet-sensitive reaction, a fraction $k_\mathrm{T}\mathrm{d}t$ potentially undergoes the triplet-sensitive reaction, and the remaining fraction, $1-k_\mathrm{S}\mathrm{d}t-k_\mathrm{T}\mathrm{d}t$, experiences no reaction. The effect of the singlet reaction is to remove the singlet component, effectively projecting the density matrix onto the remainder of the space with the projection operator, $\hat{\bar{Q}}_\mathrm{S} = \hat{E} - \hat{Q}_\mathrm{S}$, and the effect of the triplet reaction is equivalent. Thus the change in the density operator is described by

$$\hat{\rho}(t+\mathrm{d}t) = (1-k_\mathrm{S}\mathrm{d}t-k_\mathrm{T}\mathrm{d}t)\hat{\rho}(t) + k_\mathrm{S}\mathrm{d}t\,\hat{\bar{Q}}_\mathrm{S}\hat{\rho}(t)\hat{\bar{Q}}_\mathrm{S} + k_\mathrm{T}\mathrm{d}t\,\hat{\bar{Q}}_\mathrm{T}\hat{\rho}(t)\hat{\bar{Q}}_\mathrm{T} \tag{11}$$

The exact form of the reaction process is not important as we only consider the density matrix describing the radical pair spin state. After tracing out the environment any irreversible reaction will constitute a quantum measurement and so give the same result.

Since $\hat{Q}_\mathrm{S} + \hat{Q}_\mathrm{T} = \hat{E}$ we have, in the {S, T} basis, $\hat{\hat{Q}}_\mathrm{S} = \hat{Q}_\mathrm{T}$ and $\hat{\hat{Q}}_\mathrm{T} = \hat{Q}_\mathrm{S}$ so that

$$\hat{\rho}(t+\mathrm{d}t) = (1 - k_\mathrm{S}\mathrm{d}t - k_\mathrm{T}\mathrm{d}t)\hat{\rho}(t) + k_\mathrm{S}\mathrm{d}t\,\hat{Q}_\mathrm{T}\hat{\rho}(t)\hat{Q}_\mathrm{T} + k_\mathrm{T}\mathrm{d}t\,\hat{Q}_\mathrm{S}\hat{\rho}(t)\hat{Q}_\mathrm{S} \qquad (12)$$

The operators in this expression do not satisfy Equation (10) so that the trace of $\hat{\rho}(t)$ is not preserved in this approach. This is because the density matrix description does not include the increase of $k_\mathrm{S}\mathrm{d}t\,\mathrm{Tr}\left\{\hat{Q}_\mathrm{S}\hat{\rho}(t)\hat{Q}_\mathrm{S}\right\} = k_\mathrm{S}\mathrm{d}t\left\langle \hat{Q}_\mathrm{S}\right\rangle(t)$ in the fractional singlet product yield or the corresponding increase in the triplet product yield; as before, when these terms are included the sum of all populations is 1.

It might initially seem surprising that the fraction of the density matrix which could have undergone a singlet reaction, but did not, is effectively projected onto the triplet subspace. This effect is, however, well known from electron shelving experiments [22], where the non-observation of potentially-observable fluorescence acts to project the system onto the shelved state [23].

An alternative derivation of Equation (12) clarifies its origin. Considering the singlet reaction as a quantum measurement, during the interval $(t, t+dt)$ a fraction $k_\mathrm{S}\mathrm{d}t$ of the spin wavefunction of the radical pair collapses onto the $\{\mathrm{S,T}\}$ measurement basis. That is, a fraction $k_\mathrm{S}\mathrm{d}t$ of $\langle \mathrm{S}|\hat{\rho}(t)|\mathrm{S}\rangle$ reacts. Additionally, a fraction $k_\mathrm{S}\mathrm{d}t$ of $\langle \mathrm{S}|\hat{\rho}(t)|\mathrm{T}\rangle$ and of $\langle \mathrm{T}|\hat{\rho}(t)|\mathrm{S}\rangle$ disappears because the collapse of the wavefunction destroys coherent superpositions in the measurement basis. A similar argument applies for the triplet reaction. As a consequence

$$\hat{\rho}(t+\mathrm{d}t) = \hat{\rho}(t) - k_{\mathrm{S}}\mathrm{d}t\Big[\hat{Q}_{\mathrm{S}}\hat{\rho}(t)\hat{Q}_{\mathrm{S}} + \hat{Q}_{\mathrm{S}}\hat{\rho}(t)\hat{Q}_{\mathrm{T}} + \hat{Q}_{\mathrm{T}}\hat{\rho}(t)\hat{Q}_{\mathrm{S}}\Big] - k_{\mathrm{T}}\mathrm{d}t\Big[\hat{Q}_{\mathrm{T}}\hat{\rho}(t)\hat{Q}_{\mathrm{T}} + \hat{Q}_{\mathrm{S}}\hat{\rho}(t)\hat{Q}_{\mathrm{T}} + \hat{Q}_{\mathrm{T}}\hat{\rho}(t)\hat{Q}_{\mathrm{S}}\Big] \qquad (13)$$

Note that $\langle m'|\hat{Q}_m\hat{\rho}(t)\hat{Q}_n|n'\rangle = \delta_{n,n'}\delta_{m,m'}\langle m|\hat{\rho}(t)|n\rangle$. Replacing $\hat{Q}_{\mathrm{S}}$ by $\hat{E}-\hat{Q}_{\mathrm{T}}$ in the $k_{\mathrm{S}}$ term and $\hat{Q}_{\mathrm{T}}$ by $\hat{E}-\hat{Q}_{\mathrm{S}}$ in the $k_{\mathrm{T}}$ term, Equation (13) becomes

$$\hat{\rho}(t+\mathrm{d}t) = \hat{\rho}(t) - k_{\mathrm{S}}\mathrm{d}t\Big[\hat{\rho}(t) - \hat{Q}_{\mathrm{T}}\hat{\rho}(t)\hat{Q}_{\mathrm{T}}\Big] - k_{\mathrm{T}}\mathrm{d}t\Big[\hat{\rho}(t) - \hat{Q}_{\mathrm{S}}\hat{\rho}(t)\hat{Q}_{\mathrm{S}}\Big] \qquad (14)$$

which can be rearranged to give Equation (12).

Equation (12) is easily converted into differential form for easier comparison with Equations (2) and (3):

$$\frac{\mathrm{d}\hat{\rho}(t)}{\mathrm{d}t} = -\mathrm{i}\Big[\hat{H},\hat{\rho}(t)\Big] - \big(k_{\mathrm{S}}+k_{\mathrm{T}}\big)\hat{\rho}(t) + k_{\mathrm{S}}\,\hat{Q}_{\mathrm{T}}\hat{\rho}(t)\hat{Q}_{\mathrm{T}} + k_{\mathrm{T}}\,\hat{Q}_{\mathrm{S}}\hat{\rho}(t)\hat{Q}_{\mathrm{S}} \qquad (15)$$

where the coherent evolution under the Hamiltonian, neglected in Equation (12), has simply been added back in. (This is legitimate as all infinitesimal evolutions effectively commute.) Equation (15) can be rewritten in Liouville space as

$$\frac{\mathrm{d}}{\mathrm{d}t}\vec{\rho}(t) = -\hat{\hat{W}}\vec{\rho}(t); \qquad \hat{\hat{W}} = \mathrm{i}\hat{\hat{H}} + (k_{\mathrm{S}}+k_{\mathrm{T}})\hat{\hat{E}} - k_{\mathrm{S}}\,\hat{Q}_{\mathrm{T}}\otimes\tilde{\hat{Q}}_{\mathrm{T}} - k_{\mathrm{T}}\,\hat{Q}_{\mathrm{S}}\otimes\tilde{\hat{Q}}_{\mathrm{S}} \qquad (16)$$

where the tilde indicates the operator transpose. When $\hat{H}=0$, the evolution is explicitly described by

$$\exp\left(-\hat{\hat{W}}t\right)=\begin{pmatrix}\exp(-k_{\mathrm{S}}t) & 0 & 0 & 0\\ 0 & \exp(-[k_{\mathrm{S}}+k_{\mathrm{T}}]t) & 0 & 0\\ 0 & 0 & \exp(-[k_{\mathrm{S}}+k_{\mathrm{T}}]t) & 0\\ 0 & 0 & 0 & \exp(-k_{\mathrm{T}}t)\end{pmatrix} \quad (17)$$

which differs from the previous result, Equation (6), in that the $\rho_{\mathrm{ST}}$ and $\rho_{\mathrm{TS}}$ superposition terms decay at the sum of the two rates, rather than the average. This rapid decay reflects the fact that both the singlet and triplet reactions can be thought of as projective measurements in the singlet/triplet basis, and as such act to decohere superpositions. This point can be made more clearly by using operator identities to rewrite Equation (16) in the $\{|\mathrm{S}\rangle\langle\mathrm{S}|,\ |\mathrm{S}\rangle\langle\mathrm{T}|,\ |\mathrm{T}\rangle\langle\mathrm{S}|,\ |\mathrm{T}\rangle\langle\mathrm{T}|\}$ basis as

$$\hat{\hat{W}}=\hat{\hat{V}}+\tfrac{1}{2}k_{\mathrm{S}}\hat{\hat{Q}}_{\mathrm{S}}^{-}\hat{\hat{Q}}_{\mathrm{S}}^{-}+\tfrac{1}{2}k_{\mathrm{T}}\hat{\hat{Q}}_{\mathrm{T}}^{-}\hat{\hat{Q}}_{\mathrm{T}}^{-}=\hat{\hat{V}}+\tfrac{1}{2}\left(k_{\mathrm{S}}+k_{\mathrm{T}}\right)\begin{pmatrix}0&0&0&0\\0&1&0&0\\0&0&1&0\\0&0&0&0\end{pmatrix} \quad (18)$$

where the difference between $\hat{\hat{W}}$ and $\hat{\hat{V}}$ corresponds exactly to additional decoherence of $\rho_{\mathrm{ST}}$ and $\rho_{\mathrm{TS}}$.

## 4. Discussion

The calculations above indicate that the conventional phenomenological approach does not correctly predict the effects of spin-selective chemical reactions on radical pair spin dynamics. It is useful to investigate whether this has any consequences. In the following we refer to the decay of $\rho_{\mathrm{ST}}$ and $\rho_{\mathrm{TS}}$ induced by the radical recombination reactions as ‘measurement decoherence’, while ‘decoherence’ refers to any decay of coherent superposition terms, however induced, including relaxation processes that restore thermal equilibrium within the spin system. The latter arise, for example, from rotational modulation of intraradical (e.g. hyperfine) interactions or from translational modulation of interradical (e.g. exchange) interactions.

For model calculations in the absence of relaxation the two methods make different predictions, which can be related to the phenomenon usually called the quantum Zeno effect [24,25]. The quantum Zeno effect arises whenever the coherent evolution of a quantum system is repeatedly interrupted by measurements which act to project the system back onto the measurement basis; if the measurements are made sufficiently frequently then the coherent evolution can be largely suppressed. For the purposes of this effect, the key aspect of the measurement process is the decoherence of the state towards the measurement basis rather than the result of the measurement, and identical behaviour arises from other decoherence processes, such as relaxation. The quantum Zeno effect has been investigated in trapped ions [26], demonstrating the projective role of null measurements, and by NMR [27], showing an example of the effect arising from artificially induced decoherence. Although the quantum Zeno effect can appear counter intuitive, the basic principles are well established. It is also important to note that although the effect is intrinsically quantum mechanical, it in essence appears as the suppression of coherent quantum behaviour in the presence of decoherence processes, including but not limited to those arising from measurement processes.

The exact form of the quantum Zeno effect depends on the details of the system; here we describe one simple example. Consider a radical pair that begins in a pure singlet state and evolves under a singlet-triplet mixing Hamiltonian. If the reaction rate constants, $k_\mathrm{S}$ and $k_\mathrm{T}$, are small compared with the mixing rate, then the radical pair will oscillate coherently between the singlet and triplet states with a slow leakage to products. If, however, the triplet reaction rate $k_\mathrm{T}$ is much larger than the mixing rate, while the singlet rate $k_\mathrm{S}$ remains small, then the singlet-triplet mixing will be largely suppressed, and the radical pair will remain in the singlet state. Intriguingly, the rapid decay of the triplet state will suppress singlet-triplet mixing even though the radical pair never acquires significant triplet character. Similar effects have been discussed for triplet-born biradicals subject to fast singlet recombination [28].

This behaviour is illustrated in Figure 2 which shows the decay of the radical pair singlet population as a function of $k_{\mathrm{T}}$ for $k_{\mathrm{S}} = 0$. As $k_{\mathrm{T}}$ is increased from a value much smaller than the singlet-triplet interconversion frequency ($\omega$), the time-dependence changes from oscillatory, to non-oscillatory decay, to quantum Zeno effect. Closely similar time-dependence is seen for the conventional and quantum measurement approaches. The most pronounced difference occurs in the limit $k_{\mathrm{T}} >> \omega$, where the singlet population decays approximately exponentially with a rate constant $2\omega^2 / k_{\mathrm{T}}$ for the quantum measurement approach and twice as fast (rate constant, $4\omega^2 / k_{\mathrm{T}}$) for the phenomenological approach. The difference can be ascribed to the doubled decoherence rate of $\rho_{\mathrm{ST}}$ and $\rho_{\mathrm{TS}}$ in the quantum measurement case leading to a more efficient Zeno effect. Berdinskii and Yakunin have noted related aspects of the phenomenological approach [29].

Most experimental spin chemical observations are made under conditions where the radical pair recombination reactions occur at rates that are comparable to or slower than the coherent spin dynamics. An experimental test of the two approaches, i.e. Equations (3) and (16), would have to be capable of determining the rate of decoherence of the off-diagonal singlet-triplet density matrix elements. Since spin relaxation processes have qualitatively similar effects on the evolution of the density matrix, this is likely to be challenging. Ideally, spin relaxation should be negligibly slow. Although this is a common assumption, it is rarely strictly valid. Failing that, one would need to know the relaxation superoperator rather accurately so that the measurement decoherence arising from the spin-selective reactions could be unambiguously quantified. Such information is generally elusive: in most cases one cannot even be sure of the dominant relaxation mechanism let alone all the magnetic and dynamic factors that determine the rates of the relevant relaxation pathways.

We recommend that the quantum measurement approach should normally be used in future simulations of spin-selective radical pair reactions. Calculations in Liouville space are no more time-consuming than the conventional approach and require only a minor modification of the kinetics superoperator (e.g. replacing $\hat{\hat{V}}$ by $\hat{\hat{W}}$ in Equation (3)). When $k_S$ and $k_T$ are equal, or can be assumed to be so (as in the exponential model [1,30-32]), simulations using the phenomenological approach can often take advantage of the fact that the recombination superoperator is proportional to $\hat{\hat{E}}$, allowing the calculation of the coherent evolution and the kinetics to be separated. In such cases the quantum measurement approach could be significantly slower and it is not clear that the extra investment of time would always be rewarded by significantly more accurate interpretations of experimental data.

In conclusion, the measurement approach is entirely equivalent to the phenomenological approach with additional decoherence and the difference between the two results is unlikely to be dramatic and probably difficult to identify.

## Acknowledgements

We are grateful to Bela Bode, Erik Gauger, Nick Green, Hannah Hogben, Jason Lau, Kiminori Maeda, Jörg Matysik, Ulrich Steiner, Christiane Timmel and Vlatko Vedral for helpful comments and conversations.

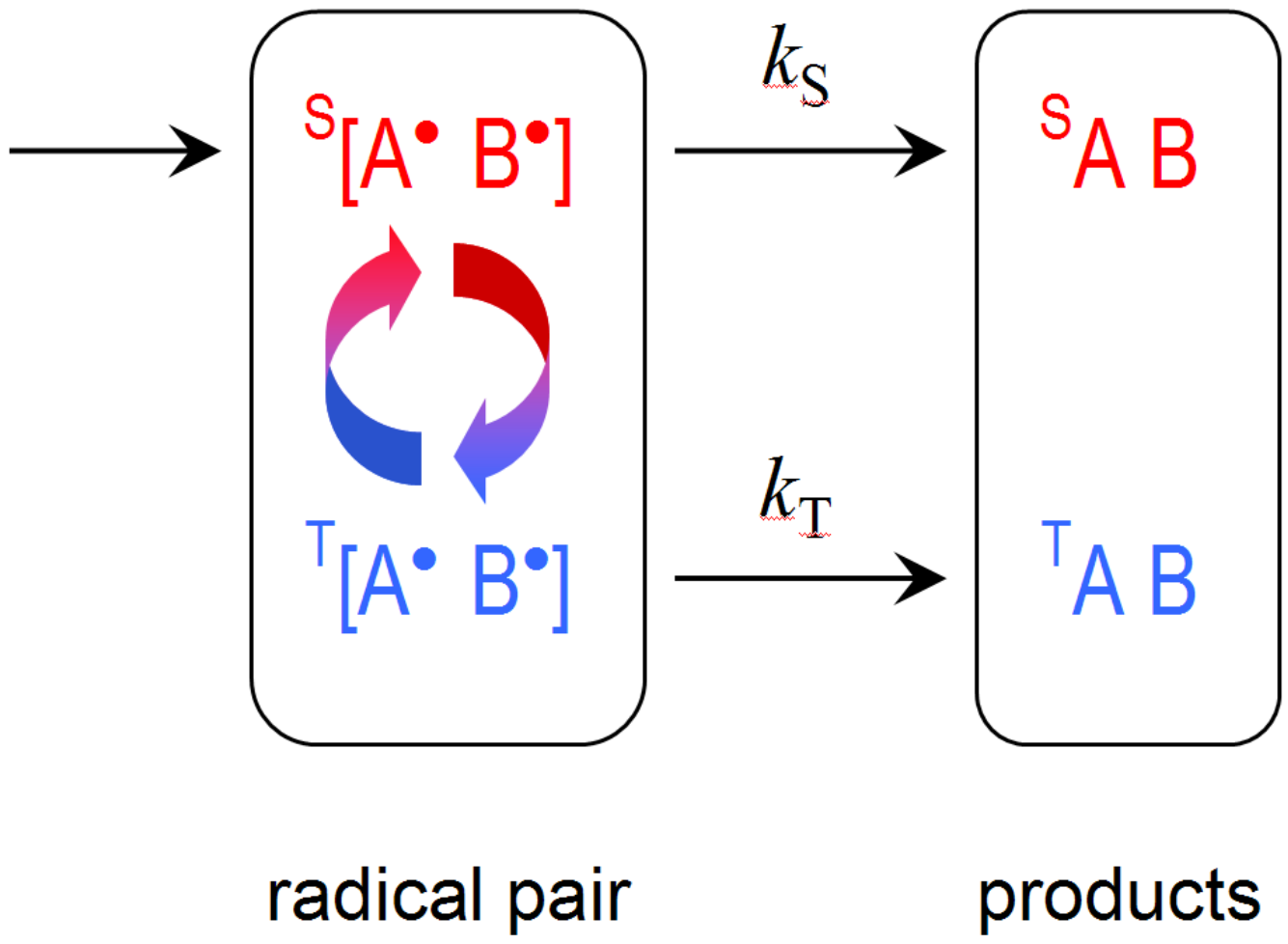

Figure 1. Simple reaction scheme for a radical pair formed initially in a singlet state. Curved and straight arrows indicate coherent singlet↔triplet interconversion and incoherent spin-selective reactions, respectively.

Figure 2A

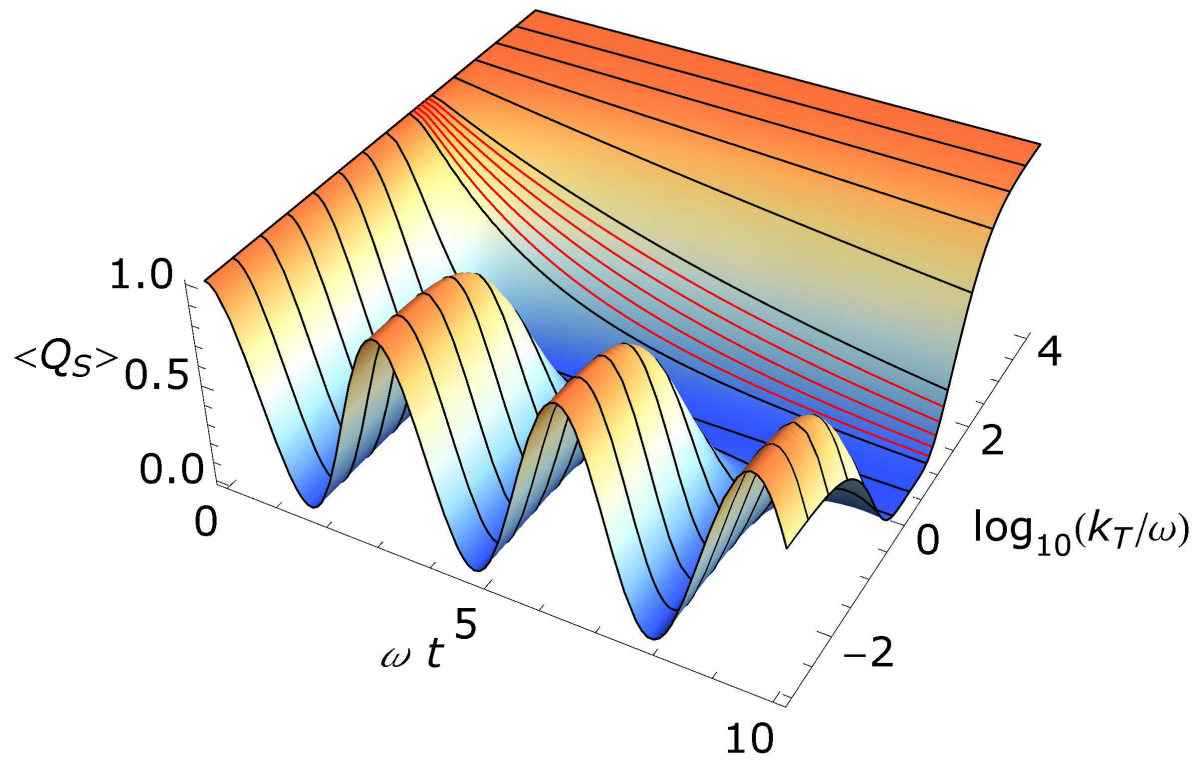

Figure 2B

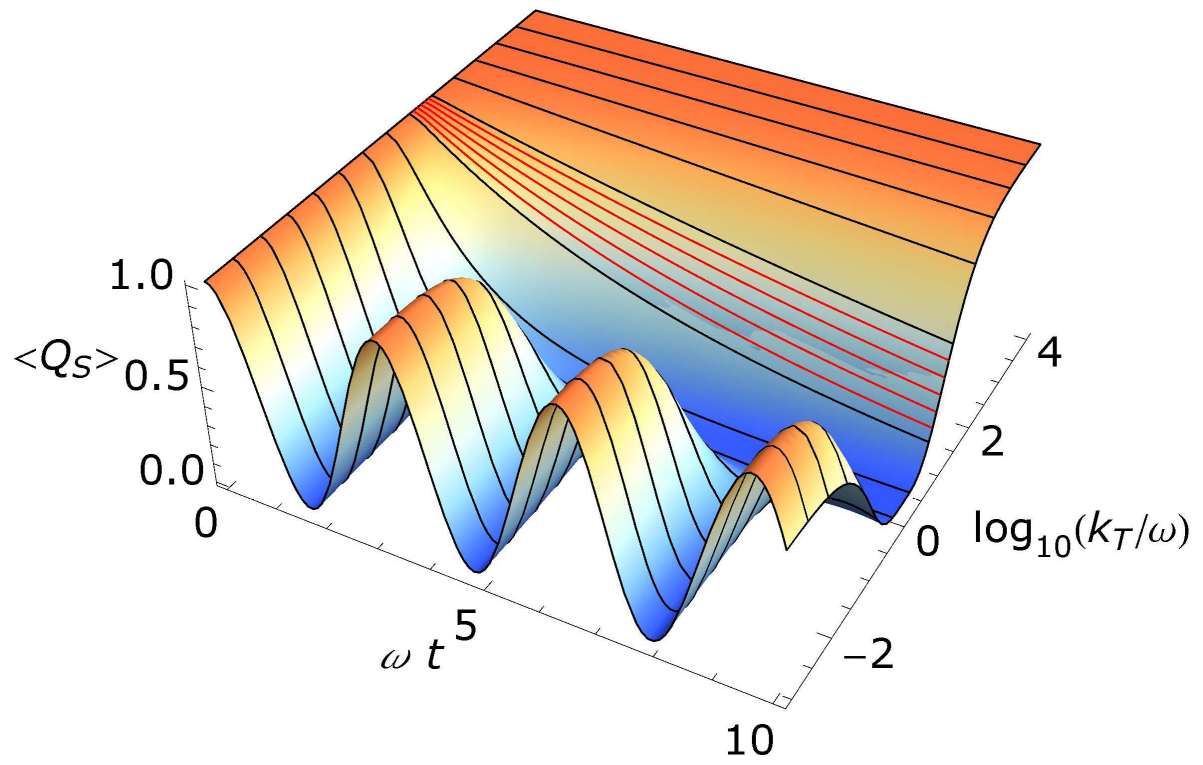

Figure 2. Simulated time-dependence of the singlet population, $\langle \hat{Q}_S \rangle$, as a function of $k_T$, for a radical pair formed at $t = 0$ in a pure singlet state. (A) the conventional phenomenological approach, Equation (3); and (B) the quantum measurement approach, Equation (16). The calculation was done using the {S,T} basis (see text) with $k_S = 0$. Both $t$ and $k_T$ are scaled by the fixed singlet-triplet mixing frequency $\omega = \langle S | \hat{H} | T \rangle$. Spin relaxation is not included. To emphasize the difference between the two approaches, the region $1.0 \leq \log_{10}(k_T / \omega) \leq 1.5$ is plotted with a red mesh.